\documentclass[
reprint,
superscriptaddress,
amsmath,amssymb,
aps,
floatfix,
]{revtex4-2}

\usepackage{graphicx,dcolumn,color,amssymb,amsmath,siunitx,setspace,tabularx,mathtools,latexsym,mathrsfs,booktabs,eucal,bm}
\usepackage{xcolor}
\usepackage[dvips]{epsfig}
\usepackage{float}
\usepackage{hyperref}
\usepackage{xr-hyper}

\begin{document}
\title{
Ca$^{2+}$-tunable mechanics and recoil in reconstituted Tcb2 networks
}
\author{Xiangting Lei}
\thanks{Correspondence to xlei45@gatech.edu}
\address{School of Chemical and Biomolecular Engineering, Georgia Institute of Technology, Atlanta, GA 30318}

\author{K. R. Prathyusha}
\address{BioFrontiers Institute, University of Colorado Boulder, Boulder, CO 80303}

\author{Carlos Floyd}
\address{Department of Chemistry and James Franck Institute, University of Chicago. Chicago, IL 60637}

\author{Jerry Honts}
\address{Department of Biology, Drake University, Des Moines, IA 50311}

\author{Saad Bhamla}
\thanks{Correspondence to saad.bhamla@colorado.edu}
\address{BioFrontiers Institute and Department of Chemical and Biological Engineering, University of Colorado Boulder, Boulder, CO 80303}

\begin{abstract}
 Tetrahymena calcium-binding protein (Tcb2) forms Ca$^{2+}$-responsive networks that exhibit contractile behavior, yet how Ca$^{2+}$ concentration controls their local mechanical response  remains poorly understood. Here, we use optical tweezers to perform active microrheology on reconstituted Tcb2 networks inside a microfluidic device that enables precise control of Ca$^{2+}$ concentration, allowing systematic tuning of network structure and mechanics. We find that increasing Ca$^{2+}$ from 1 to 100~mM enhances the effective stiffness by nearly two orders of magnitude, from $\sim$1$\times$10$^{-3}$ to $\sim$7$\times$10$^{-2}$~pN/nm, corresponding to a transition from a viscosity-dominated to a more elastic and mechanically robust network. Recoil assays further reveal rapid release of stored elastic energy following deformation. 
  After two orthogonal pulls, the bead recoils along the diagonal rather than retracing the loading path, indicating that stresses from different directions combine to produce a resultant restoring response.  These results establish Tcb2 networks as a minimal, tunable system for probing chemomechanical coupling and viscoelasticity in Ca$^{2+}$-regulated protein networks.
\end{abstract}

\maketitle

\section{Introduction}

Protein-based fibrous materials occupy a distinctive position within the broader class of soft materials~\cite{Schiller2024-hj}. Major examples include cytoskeletal networks formed by actin filaments, microtubules, and intermediate filaments~\cite{Fletcher2010-en}, extracellular fibrous matrices like collagen and fibrin~\cite{Frantz2010-cq}, and reconstituted protein hydrogels~\cite{Banta2010-si}.  Proteins can undergo conformational changes and participate in multivalent interactions and reversible binding events that directly couple biochemical cues to mechanical responses. As a result protein assemblies can form dynamic networks whose structure and mechanics are highly tunable, often over biologically relevant length and time scales. 
Understanding how molecular-scale interactions within such networks give rise to emergent mechanical behavior remains a key challenge for the development of bioinspired soft materials with  tunable mechanical properties~\cite{Schiller2024-hj,DeFrates2018-ob}.

The mechanical behavior of protein-based fibrous materials typically arises from interconnected assemblies of filament networks, whose response is governed by connectivity, filament flexibility, and inter-filament dynamics~\cite{Broedersz2014-ad,Lieleg2010-tp}. For instance, in cytoskeletal systems, 
transient crosslinking between filaments enables viscoelastic stress relaxation, while molecular motors acting on inherently polar filaments generate internally driven stresses that can profoundly alter network organization and mechanical response, giving rise to nonequilibrium phenomena such as contractility \cite{Murrell2015-cj,Lenz2012-co}, and active stiffening~\cite{Mizuno2007-he}.
The balance between elastic energy storage and viscous dissipation is therefore governed not by the properties of individual filaments, but by how they are connected and how long those connections persist.  Even though these effects are essential for the proper functioning of the cell, they introduce internally generated stresses~\cite{Koenderink2009-oz,Mizuno2007-he}, polarity-dependent activity \cite{Murrell2015-cj,Vicente-Manzanares2009-iu}, continuous structural remodeling \cite{Lieleg2011-cr,Kohler2012-qx}, and persistent nonequilibrium dynamics that complicate the interpretation of emergent material behavior.

Motor-free reconstituted protein networks have emerged as useful model systems for probing the physical origins of network mechanics. By eliminating motor driven activity, these systems allow the roles of filament architecture, crosslink density, and crosslink dynamics in governing viscoelastic behavior to be isolated~\cite{Broedersz2014-ad,mackintosh1995elasticity}. 
For instance, studies on crosslinked actin networks have shown that crosslinker concentration and unbinding rate govern the transition from fluid-like to solid-like behavior and give rise to strain stiffening~\cite{Tharmann2007-yz,Gardel2004-science,Gardel2006-pnas}, while dynamic crosslink models demonstrated that crosslink lifetime sets a characteristic stress relaxation timescale of the network~\cite{Lieleg2008-hl,Lieleg2009-bj}. More generally, deformation can reorganize the structure of dynamically connected networks and alter their rheological response\cite{schmoller2009structural,prathyusha2013shear,schmoller2010cyclic}
Intermediate filament, fibrin, and collagen networks further showed how filament compliance, hierarchical assembly, and inter-filament interactions collectively tune nonlinear elasticity and mechanical resilience across a wide range of length and time scales~\cite{Storm2005-ur}. 
Motor-independent force generation can also arise through filament depolymerization~\cite{grishchuk2005force} or the entropic expansion of diffusible crosslinkers confined between overlapping filaments~\cite{lansky2015diffusible}. However, directly regulating network connectivity and mechanics through a specific biochemical control parameter remains challenging.


 
Network connectivity can also be regulated through ionic interactions. In actin networks, Mg$^{2+}$ counterions form crossbridges between filaments, promoting bundling and altering network mechanics~\cite{gurmessa2019counterion}, while changes in Mg$^{2+}$ concentration can induce motor free network contraction~\cite{ricketts2020triggering}. However, these interactions arise primarily from the polyelectrolyte nature of the filaments rather than a specific ion binding mechanism. This motivates systems in which connectivity is governed by specific, controllable interactions. Tetrahymena calcium binding protein 2 (Tcb2) provides such a system: it is a Ca$^{2+}$ sensitive EF hand protein that undergoes conformational changes upon Ca$^{2+}$ binding and forms contractile networks in the presence of Ca$^{2+}$. Recently, we demonstrated that Tcb2 forms Ca$^{2+}$ triggered contractile networks \textit{in vitro}, with assembly, growth, and spatiotemporal dynamics controlled through optical Ca$^{2+}$ release~\cite{Lei2025-dq}.


Despite this control over network assembly, how Ca$^{2+}$ concentration governs the local mechanical properties of Tcb2 networks, a single component system that exhibits Ca$^{2+}$ induced contractile behavior in the absence of molecular motors, remains largely unexplored. In this work, we aim to understand the relationship between Ca$^{2+}$ dependent network organization and  mechanical properties. To address this question, we integrate multichannel microfluidics for precise control of Ca$^{2+}$ concentration~\cite{Colin2020-nr}, confocal imaging to characterize network structure, and optical tweezer microrheology to measure the force response to controlled probe displacement~\cite{Robertson-Anderson2018-gn,Lehmann2020-hv}. From these measurements, we quantify the Ca$^{2+}$ dependent evolution of effective network stiffness and probe the elastic response through recoil measurements following deformation. These results demonstrate that calcium can act as a sensitive tune for filamentous network mechanics.

\section{Results}
\subsection{Microfluidic assay for Ca$^{2+}$-responsive Tcb2 networks}

\begin{figure*}[!]
\centering
\includegraphics[width=1.05\textwidth]{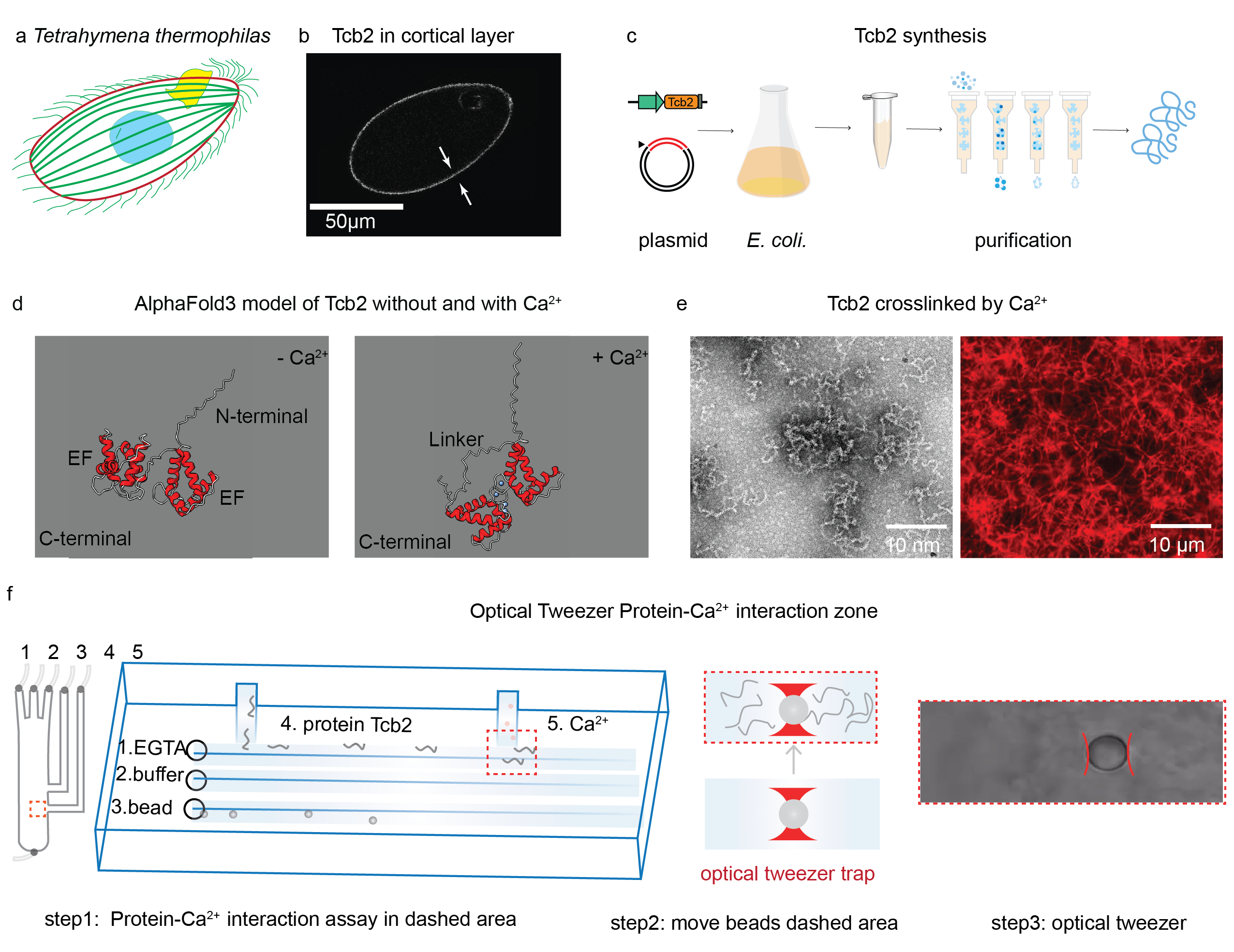}
\caption{
    \textbf{Tcb2 structure, mechanics and microfluidic chip design. }
    (a) \textit{Tetrahymena Thermophila} uses the protein Tcb2 for rapid motion.  
    (b) Confocal microscopy image of \textit{Tetrahymena} with labeled Tcb2, showing its localization at the cortical layer between the membrane and the cytoskeleton.  
    (c) Workflow of Tcb2 protein synthesis and purification. See Methods for details.  
    (d) Alphafold3 model of Tcb2 protein structure change without and with Ca$^{2+}$. Labeled are the calcium-binding domains (EF hands).  Adapted from ~\citenum{Lei2025-dq}.
    (e) Transmission electron microscopy (TEM) image of synthesized Tcb2 showing filaments at the nanometer scale at low Ca$^{2+}$ concentrations, and scanning electron microscopy (SEM) image showing cross-linked networks at the micron scale at high Ca$^{2+}$ concentrations. Adapted from ~\citenum{Lei2025-dq}.
    (f) Microfluidic setup for introducing and manipulating optically trapped beads within Tcb2 networks.}
\label{M-figure1}              
\end{figure*}  
\textit{Tetrahymena thermophila} is a ciliated eukaryotic protist (Fig.~\ref{M-figure1}a) that uses Tcb2, a calcium-binding protein forming a dense mesh between the membrane and cytoskeleton, to generate rapid contractions (Fig.~\ref{M-figure1}b).
In previous work, we established a protocol to express and purify Tcb2 \textit{in vitro} (Fig.~\ref{M-figure1}c; see Methods) \cite{Lei2025-dq}. Tcb2 monomer contains two calcium-binding domains, each with two EF-hands. Upon binding to Ca$^{2+}$, the N- and C-terminal domains move closer together, a conformational change predicted by AlphaFold3 and consistent with NMR data \cite{Kilpatrick2016-rv} (Fig.~\ref{M-figure1}d). This rearrangement drives crosslinking and network assembly at larger scales: at 1 $\mu$M Ca$^{2+}$, Tcb2 forms short filaments of $\sim$200 nm (TEM), while at 1 mM Ca$^{2+}$, these filaments assemble into a homogeneous crosslinked network (SEM, Fig.~\ref{M-figure1}e)~\cite{Lei2025-dq,Chandrasekharan2025-sx}.

Here, we build on previous work to characterize the mechanical properties of the Tcb2 network. Conventional bulk rheometry is not suitable because the protein is highly sensitive to Ca$^{2+}$ and tends to form local aggregates: it quickly phase-separates into filamentous and cross-linked regions due to the fast binding rate ($\sim$50\,mM$^{-1}$\,s$^{-1}$) and the low Ca$^{2+}$ diffusivity ($\sim$300\,$\mu$m$^2$/s). Direct bulk viscosity measurements give $\sim$1.2\,cP at 1\,mM and $\sim$0.9\,cP at 100\,mM Ca$^{2+}$. Bulk rheometry therefore averages over the buffer-rich regions and cannot resolve the local response of cross-linked subregions probed by optical tweezers (Fig.~S1). To address this, we combine optical-tweezer microrheology with a custom microfluidic chip that provides precise control over the Ca$^{2+}$ concentration at the microscale.

We introduce purified Tcb2 proteins into a custom-made microfluidic device (Fig.~\ref{M-figure1}f) with five inlet channels.
Three parallel channels deliver EGTA (a calcium chelator that prevents aggregation), Tris buffer (to maintain pH 7.5), and 4.98 $\mu$m polystyrene beads. Two perpendicular channels introduce Tcb2 and Ca$^{2+}$ separately, which meet at a designated interaction zone (red dashed box, Fig.~\ref{M-figure1}f), where they mix uniformly and form a crosslinked network.
By varying the Ca$^{2+}$ concentration in the fifth channel, we can probe the protein and its microrheological properties across a broad range of Ca$^{2+}$ concentrations from 1 to 100~mM.

\subsection{
Ca$^{2+}$ tunes the structure and stiffness of Tcb2 networks}

\begin{figure*}[t]
    \centering
    \includegraphics[width=0.90\textwidth]{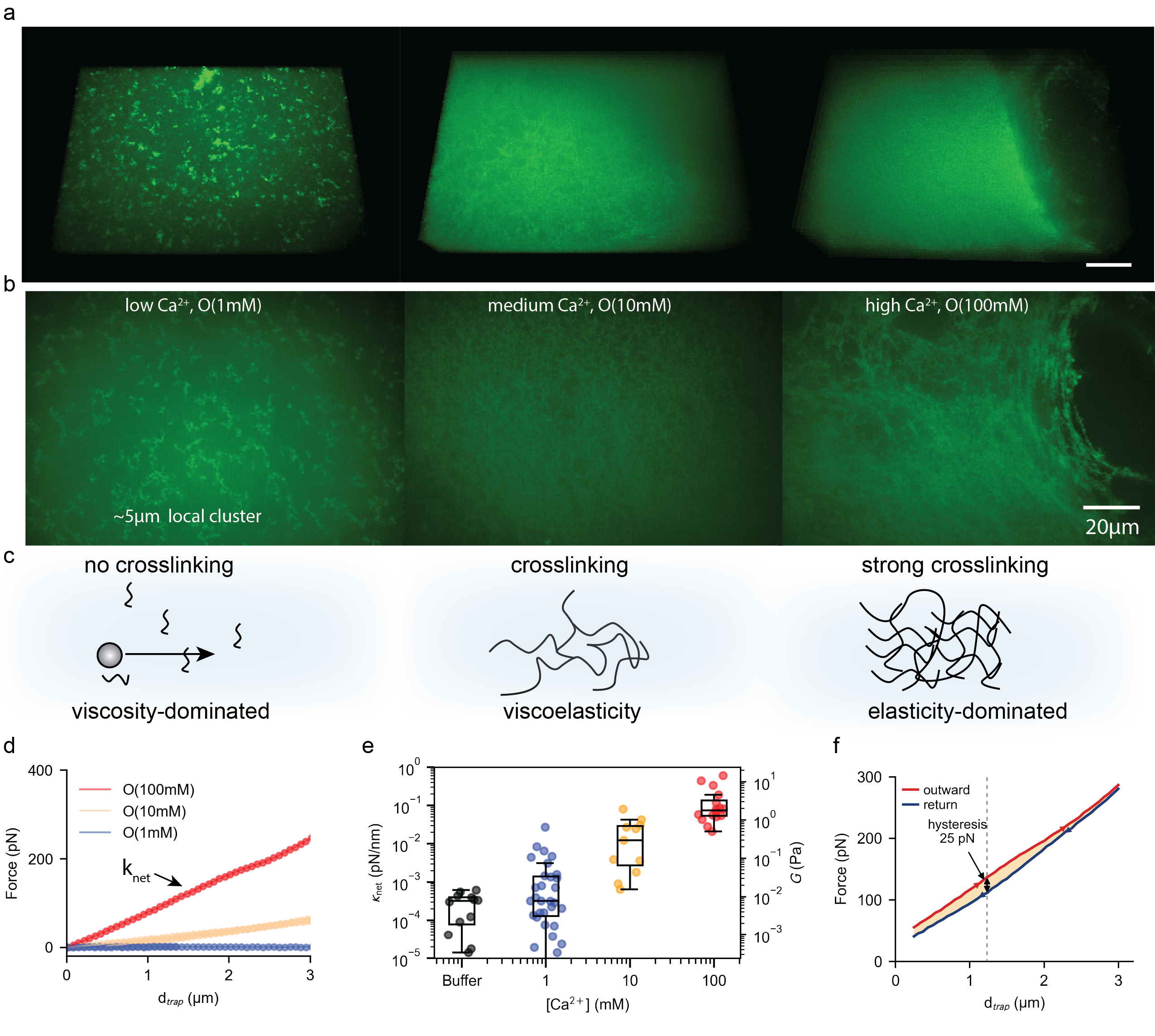}
    \caption{
    \textbf{Microscopic structure and mechanical properties of Tcb2 networks as a function of Ca$^{2+}$ concentration.}
    (a) 3D confocal reconstructions and (b) corresponding 2D confocal cross-sections of Tcb2 networks at 1, 10, and 100 mM Ca$^{2+}$.
    (c) Schematic of the optical tweezer protocol: a trapped bead is moved at 1 $\mu$m/s through the network.
    (d)Force–displacement curves at 1, 10, and 100 mM Ca$^{2+}$.
    (e) Effective stiffness $\kappa_\mathrm{net}$ (pN/nm) extracted from the linear region of the force–displacement curves in (d).
    (f) Hysteresis effect at 100 mM Ca$^{2+}$.Outward and return branches of a single out-and-back cycle from one bead, one of 13 cycles recorded.}
    \label{M-figure2}
\end{figure*}

To understand how Ca$^{2+}$ concentration controls both network structure and mechanical response, we first characterize the morphology and then quantify the microrheological properties of Tcb2 networks. 

We use confocal imaging to map the protein network morphology at different Ca$^{2+}$ concentrations (Fig.~\ref{M-figure2}a,b). The panels in (a) show reconstructed 3D images, and the panels in (b) show high-contrast 2D cross-sections of Tcb2 networks across three Ca$^{2+}$ concentrations. Consistent with the electron microscopy in Fig.~\ref{M-figure1}e, we observe that Ca$^{2+}$ addition drives the assembly of Tcb2 monomers into filamentous structures spanning the nano- to micron scale.
At 1 mM calcium, Tcb2 forms small clusters of approximately 5 $\mu$m in diameter (Fig.~\ref{M-figure2}a; left) (SI Movie section I).
At 10 mM,  the assembly transitions into a connected network~(Fig.~\ref{M-figure2}a; middle). 
At 100\,mM, the assembly undergoes a distinct transition, forming dense, contracted structures with boundary-initiated rupture contraction (Movie S1, Section I).
Thus, as the Ca$^{2+}$ concentration is varied, the network undergoes substantial changes in morphology and local density, consistent with a transition from a loosely connected meshwork to a dense, highly connected network.

To quantify how these structural changes translate into mechanical response, we next perform microrheological measurements using optically trapped particles. 
Beads are trapped using optical tweezers and positioned in the interaction zone (Fig.~\ref{M-figure1}f, right panel) to locally probe the material.
In each experiment, beads are moved at a fixed speed of 1~$\mu$m/s (Fig.~\ref{M-figure2}c) and we measure the force over time $F(t)$ required to maintain this displacement constant until the displacement $d(t)$ reaches 10~$\mu$m. 
The measured force corresponds to the restoring force exerted by the optical trap, which balances the resistance experienced by the bead as it moves through the network, and therefore reflects the effective mechanical response of the surrounding material. 
Thus this protocol allows us to distinguish between elastic and viscous responses by examining how forces scale with displacement and velocity. 
For purely elastic protein networks, forces increase linearly with displacement but remain constant with respect to velocity. 
Conversely, in viscous protein networks, forces increase linearly with velocity but are independent of displacement.
We measured that the slope $k$ from our force--displacement curves is constant across both pull displacement (5-20\,$\mu$m, Fig.~S2)
and pulling velocity (1, 2, 5\,$\mu$m/s, Fig.~S3) at all three Ca$^{2+}$ concentrations, suggesting that $k$ reflects a linear elastic contribution to the network response more than a velocity-dependent drag. At 100~mM the 20,$\mu$m condition exceeds the trap's maximum holding force (Fig.~S4).

The force-displacement curves for different calcium concentrations are shown in Fig.~\ref{M-figure2}d. At low Ca$^{2+}$ concentration (1 mM), the force remains close to zero across the entire displacement range, indicating minimal resistance to deformation and a weakly connected network. As the concentration increases to 10 mM, the force rises gradually with displacement,  reflecting the emergence of a finite resistance as filament associations begin to span larger length scales. At 100 mM Ca$^{2+}$, the force increases steeply and approximately linearly with displacement, signaling a substantial gain in mechanical resistance.

Because the force–displacement response in Fig.~\ref{M-figure2}d is approximately linear at small deformations, we extract an effective stiffness $k = \partial F / \partial d$ 
from the slope of the curves, as summarized in Fig.~\ref{M-figure2}e.
At 1 mM Ca$^{2+}$, this modulus  k is negligible around $1\times10^{-3}$\,pN/nm , comparable to the buffer solution, indicating that the material behaves predominantly as a viscous fluid.
At 10 mM, $k$ rises to $8\times10^{-3}$\,pN/nm, reflecting a finite elastic resistance to deformation. At 100 mM, the stiffness is $7\times10^{-2}$\,pN/nm, around two orders of magnitude larger than that at 1~mM, confirming that denser cross-linking yields a much stiffer material. 
We calculated the shear modulus via the generalized Stokes–Einstein relation, these values correspond to $G \approx 0.024$, 0.19, and 1.7 Pa (Table~S1).

The continuous increase suggests a progressive enhancement of network connectivity with increasing Ca$^{2+}$, rather than an abrupt transition. 
The intermediate regime shows the emergence of a finite stiffness, consistent with the formation of a system-spanning network, while the broader distribution at higher concentrations indicates increased heterogeneity in local mechanical response, likely reflecting spatial variations in network structure. Altogether, these results show that the Ca$^{2+}$ concentration can be used to tune both the structure and mechanics of Tcb2 networks.

At 100\,mM Ca$^{2+}$ we observe hysteresis in the force--displacement curves under back-and-forth bead motion (Fig.~\ref{M-figure2}f): the outward and
return paths do not overlap. In the bead shown the return branch lies below
the outward branch in all $13$ consecutive cycles, by up to $25$\,pN in the
cycle plotted. We reproduced this on $n = 3$ independent beads. The difference falls to 6.0\,pN at 10\,mM and -2.0\,pN at 1\,mM (Fig.~S5). 

Such history dependence is  characteristic of transiently cross lined polymer networks, in which bonds rupture and reform at new positions during deformation ~\cite{Lieleg2008-hl}. The effect is nonetheless
small comparing to the values we moved the beads, so the network is predominantly elastic with a minor dissipative
component. It thus exhibits both elastic energy storage and dissipative relaxation, the former probed directly by the recoil experiments in the next section. These results demonstrate that calcium can act as a sensitive tune for filamentous network mechanics.

\subsection{Recoil experiments reveal calcium-dependent elastic recovery}

\begin{figure*}[!]
\centering
    \includegraphics[width= 0.9\textwidth]{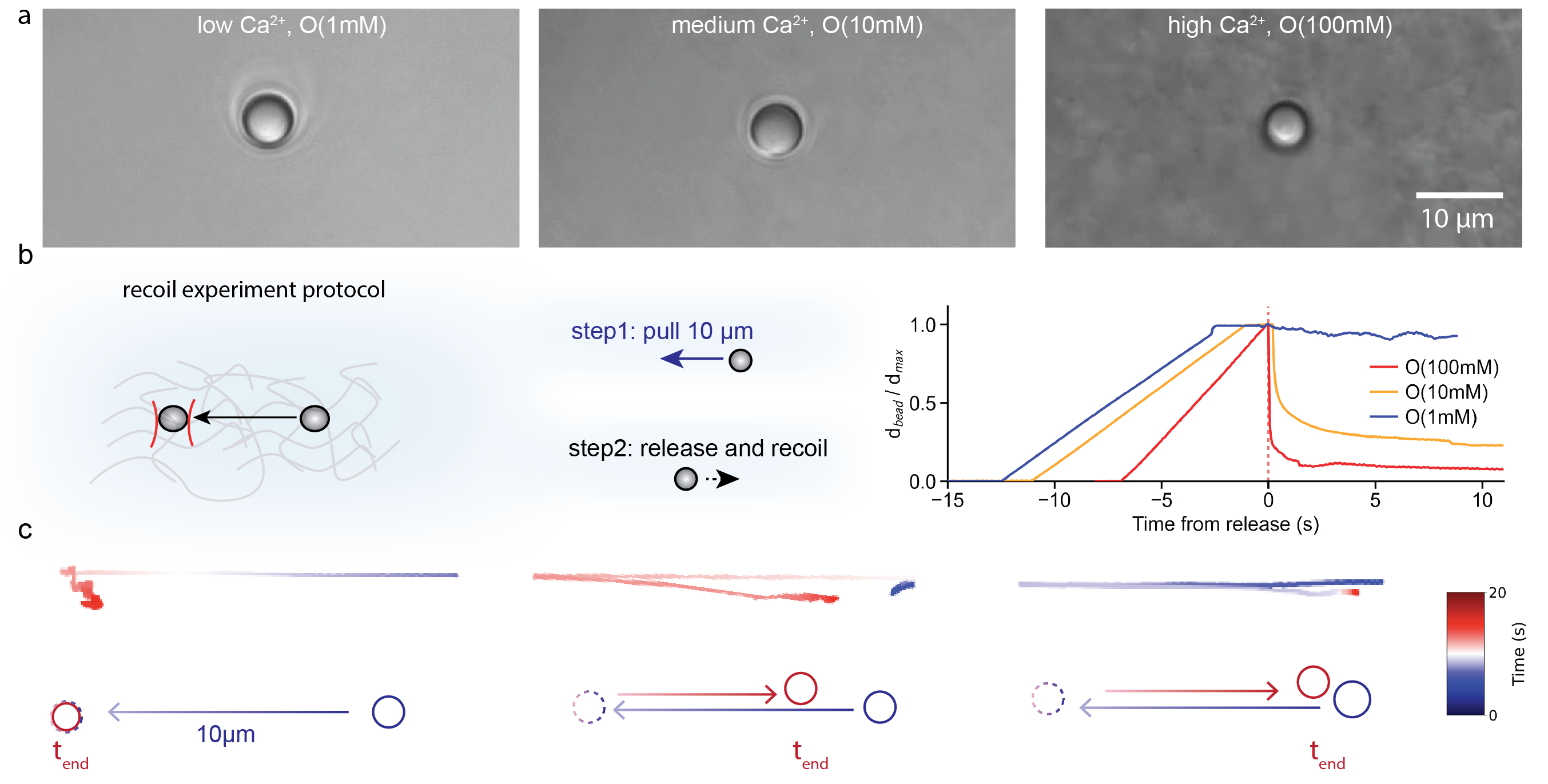}
    \caption{\textbf{Recoil experiments to probe viscoelasticity}.
    (a) Images of beads in quiescent protein network at 1, 10 and 100 mM Ca$^{2+}$ concentrations.
    (b) Schematic of the experimental protocol: a bead is displaced 10 $\mu$m at 1 $\mu$m/s by the optical trap and then released. At 100 mM the network force exceeds the trap's maximum ($\sim$1000 pN) and the bead escapes around 5 $\mu$m. Bead displacement normalized to its value at release. More detailed in figure S2.
    (c) Time-colored bead trajectories during recoil at 1, 10, 100 mM Ca$^{2+}$ concentration.
    }
    \label{M-figure3}
\end{figure*} 

Hysteresis reveals that the network dissipates energy and responds in a history-dependent manner, but it does not directly show how much elastic energy is stored or how quickly it is released. 
To probe these properties, we use a ``recoil'' assay that measures the elastic recovery of beads after displacement by the optical trap. 
First, we image the beads in their quiescent state embedded in the protein network at different Ca$^{2+}$ concentrations (Fig.~\ref{M-figure3}a).
We first image the beads in their quiescent
state, embedded in the protein network at different Ca$^{2+}$ concentrations (Fig.~\ref{M-figure3}a). We then trap each bead with the optical tweezers,
translate the trap by 10\,$\mu$m at 1\,$\mu$m/s, switch the trap off, and track the subsequent recoil (Fig.~\ref{M-figure3}b,c). The bead follows the
trap to 10.1\,$\mu$m at 1\,mM and 8.5\,$\mu$m at 10\,mM; at 100\,mM the restoring force of the network reaches ${\sim}2.5$\,nN, exceeding the maximum
force of the trap, and the bead escapes at ${\sim}5\,\mu$m.

A purely elastic network would fully recoil upon release, whereas a purely viscous one would show no recoil at all~\cite{Robertson-Anderson2018-gn}. Elastic recovery increases
systematically with Ca$^{2+}$: the bead recovers 7\% of its displacement at 1\,mM, 79\% at 10\,mM and 93\% at 100\,mM. This progression therefore directly measures the elastic fraction of the network response. More recoil data are quantified in Fig.~S6 (see also Movie S1, Section II). Together with the stiffness measurements, these results establish a calcium-mediated shift from a predominantly fluid-like response to a predominantly elastic one, with
a small dissipative component that persists across the whole range.

%


\subsection{Diagonal recoil after sequential two-dimensional L-shaped deformation at 100 mM Ca$^{2+}$ concentration}
\begin{figure*}[!]
\centering
    \includegraphics[width= 0.9\textwidth]{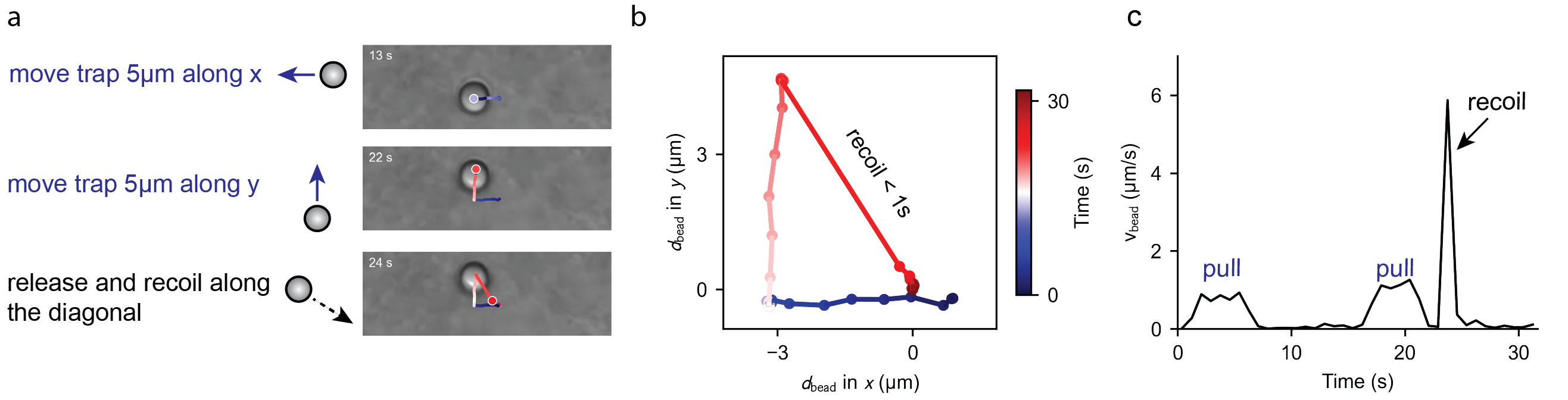}
    \caption{
    \textbf{Two-dimensional sequential recoil experiments at 100 mM Ca$^{2+}$}.
    (a) The trap is translated 5\,$\mu$m along $x$, held, translated 5\,$\mu$m along $y$, and then switched off; the bead recoils along the diagonal rather than retracing the path. 
    (b) Bead displacement in $x$--$y$, coloured by time.
    (c) Bead speed over time.
    }
 \label{M-figure4}
\end{figure*}

To probe the elastic response of Tcb2 networks to sequential deformations, we extended our analysis from single-direction to multi-directional recoil tests. At 100 mM Ca$^{2+}$, we displaced an optically trapped bead by 5 $\mu$m along the x-axis, then by 5 $\mu$m along the y-axis, and released it from the trap (Fig.~\ref{M-figure4}a ; see also Movie S1, Section II)

The trajectory is shown in Fig.~\ref{M-figure4}b. The
two slow, orthogonal displacements are followed by a single rapid recoil, and the bead does not retrace the L-shaped path defined by the pulls. Instead, it returns along the a straight line along the diagonal in less than a second. 
The peak recoil velocity 6\,$\mu$m/s exceeds the
1\,$\mu$m/s pulling velocity by roughly a factor of five
(Fig.~\ref{M-figure4}c), indicating the rapid release of stored elastic stress
on removal of the trap, while the incomplete recovery along $x$ indicates that
part of the earlier deformation was dissipated by internal rearrangement during
the interval before release. 

Taken together, these observations indicate that the two orthogonal deformations are stored as a single elastic configuration rather than as two independent ones: on release the bead follows one straight path, as expected if the network relaxes toward a single stress-free state, rather than reversing the two pulls in sequence.

\section{Discussion}
Here we investigated the chemomechanical properties of a reconstituted protein network using a bottom-up approach within a microfluidic platform. The system is based on Tetrahymena calcium-binding protein (Tcb2), which exhibits Ca$^{2+}$-dependent assembly and contractile behavior. 
Using optical tweezer microrheology, we showed that Ca$^{2+}$ concentration tunes the effective stiffness over a wide range, from $\sim$1$\times$10$^{-3}$  at 1\,mM to $\sim$7$\times$10$^{-2}$ ~pN/nm at 100\,mM Ca$^{2+}$, placing Tcb2 within the range spanned by single motor proteins -- comparable to the binding stiffness of dynein ($\sim$0.05\,pN/nm) though well below myosin~II ($\sim$1.8\,pN/nm) -- despite being a passive, single-component system (Table~S2). This tunability is accompanied by a transition from a weakly connected, fluid-like response to a more elastic and mechanically robust network. Complementary recoil experiments demonstrate that the network is capable of storing and releasing mechanical energy. Also the diagonal recoil shows the network could relax in a coordinated way, with stresses from different directions combining into a single response.

These results establish Tcb2 as a minimal, single-protein system in which network mechanics and directional coupling can both be tuned through a single chemical input. The combination of motor-free assembly, Ca$^{2+}$-dependent stiffness control, and directional recoil distinguishes it from cytoskeletal and motor-driven systems and provides a quantitative reference point for future work on Ca$^{2+}$-responsive biopolymer networks.

\section{Materials and Methods}

\subsection{Protein expression and purification}

Protein expression and purification followed our previously published protocol~\cite{Lei2025-dq}. Briefly, the Tcb2 gene~\cite{ncbi-Tcb2} was cloned into pJ411 with T7 promoter and transformed into \textit{E.~coli} BL21(Fig.~S7). Cells were grown in LB or SB with kanamycin, induced with 1mM IPTG, and left to express overnight at 18$^\circ$C. After lysis with B-PER and clarification (15000~g, 15 min), the pellet was extracted in 4 M urea, 0.25 mM EGTA, and 25 mM Tris–HCl (pH 7.5). The supernatant was loaded onto a 25 mL Q-Sepharose column and eluted with a NaCl step gradient in 4 M urea.

Before each experiment, urea was removed by overnight dialysis (3.5–5 kDa MWCO, Spectra‑Por Float‑A‑Lyzer G2) against 25 mM Tris–HCl (pH 7.5) with 1 mM EGTA. The protein was concentrated to $\sim$ 30 mg/mL using a centrifugal filter (Amicon Ultra,UFC500324), and the final concentration was measured by absorbance at 280 nm.

\subsection{Sample preparation and microfluidics}

Experiments were performed in a five-channel microfluidic chip (LUMICKS, SKU:~9181) integrated with the C-Trap optical-tweezer platform. Three parallel channels were loaded with 200~$\mu$L of buffer containing 40~mM Tris and 1~mM EGTA, and 500~$\mu$L beads, 4.0--4.9~$\mu$m in diameter, at 0.005\%~w/v in PBS with 3~mM sodium azide. Two perpendicular channels were loaded with 200~$\mu$L of 30~mg/mL Tcb2 and 200~$\mu$L of Ca$^{2+}$ solution with 1, 10, or 100~mM concentration.

\subsection{Optical tweezer setup}
Microrheology measurements were performed on a C-Trap confocal-fluorescence optical-tweezer system (LUMICKS) operated at 30\% of maximum laser power. Force was recorded at both high frequency (Force HF, 78{,}125~Hz) and low frequency (Force LF, 15~Hz). Trap position was recorded at 78{,}125~Hz, and bright-field images at 15~fps.  Bead trajectories from bright-field video were analyzed using Co-Tracker with a manual seed point and a 50-pixel grid.

All experiments used a bead at default 1~$\mu$m/s, with force and position recorded continuously. Four protocols were used. For force--displacement measurements, the bead was pulled through the network until the displacement reached $d = 10$~$\mu$m, and the effective stiffness $k = \partial F/\partial d$ was extracted from a linear fit at small $d$. These experiments were performed at 1, 10, and 100~mM Ca$^{2+}$. For hysteresis experiment,  the bead was driven back and forth with a 10~$\mu$m amplitude for 20 cycles at 100~mM Ca$^{2+}$, For single-direction recoil experiments, the bead was displaced by 5~$\mu$m, the trap was released, and its subsequent position was tracked until it reached steady state. These experiments were performed at 1, 10, and 100~mM Ca$^{2+}$. For sequential two-dimensional recoil experiments, the bead was first displaced by 10~$\mu$m along the $x$-axis, then by 10~$\mu$m along the $y$-axis, and then released. These experiments were performed at 100~mM Ca$^{2+}$.

\subsection{Confocal imaging}
Three‑dimensional imaging was carried out on a Dragonfly 400 spinning‑disk confocal microscope (Andor Technology). Samples containing 250 $\mu$M Tcb2 and 100µM EGTA were prepared in reaction buffer, and 1µL of the solution was sealed in a fixed‑height chamber for imaging.  Image stacks were acquired at 30 fps using three immersion objectives—40× , 60× , and 100× —under Ca$^{2+}$ concentrations of 1, 10, and 100 mM. For each field of view, a 50 $\mu$m z‑range was sampled at 1$\mu$m optical steps, yielding 50 sections per stack. ImageJ was used for further analysis and movie generation.\\

\subsection{Bulk rheology} 

Bulk rheology was characterised with a microfluidic rheometer RheoSense maintained at $20~^\circ\mathrm{C}$. After calibration with de‑ionised water ($\eta = 1.0~\mathrm{cP}$), protein solutions were loaded into a clean, temperature‑equilibrated microchip and subjected to oscillatory frequency sweeps from $0.1$ to $100~\mathrm{rad}~\mathrm{s}^{-1}$ at a constant strain amplitude of $1\%$. Each sample was measured three to four times; between runs the chip and flow path were flushed with DI water, rinsed with $1\%$ (v/v) Aqut cleaning solution, re‑flushed with DI water, and the instrument recalibrated.

\section{Acknowledgment}
We acknowledge the Marine Biological Laboratory Whitman Summer Investigator Program and the Physiology Course for the use of microscopy equipment and optical tweezers. S.B. acknowledges support from NSF award MCB-2313724, National Institutes of Health (NIH) award R35GM142588 and Schmidt Sciences, LLC. 

\bibliographystyle{unsrt}
\bibliography{proteinnetwork}


\end{document}


\title{Supplemental Material for: \\
Ca$^{2+}$-tunable mechanics and recoil in reconstituted Tcb2 networks}

\author{Xiangting Lei}
\thanks{Correspondence to xlei45@gatech.edu}
\address{School of Chemical and Biomolecular Engineering, Georgia Institute of Technology, Atlanta, GA 30318}

\author{K. R. Prathyusha}

\address{BioFrontiers Institute, University of Colorado Boulder, Boulder, CO 80303}
\author{Carlos Floyd}
\address{Department of Chemistry and James Franck Institute, University of Chicago, Chicago, IL 60637}

\author{Jerry Honts}
\address{Department of Biology, Drake University, Des Moines, IA 50311}

\author{Saad Bhamla}
\thanks{Correspondence to saadb@chbe.gatech.edu}
\address{BioFrontiers Institute and Department of Chemical and Biological Engineering, University of Colorado Boulder, Boulder, CO 80303}

\maketitle
\onecolumngrid
\tableofcontents
\newpage

\section{Bulk viscosity measurements}

\begin{figure}[h!]
\centering
    \includegraphics[width=0.4\textwidth]{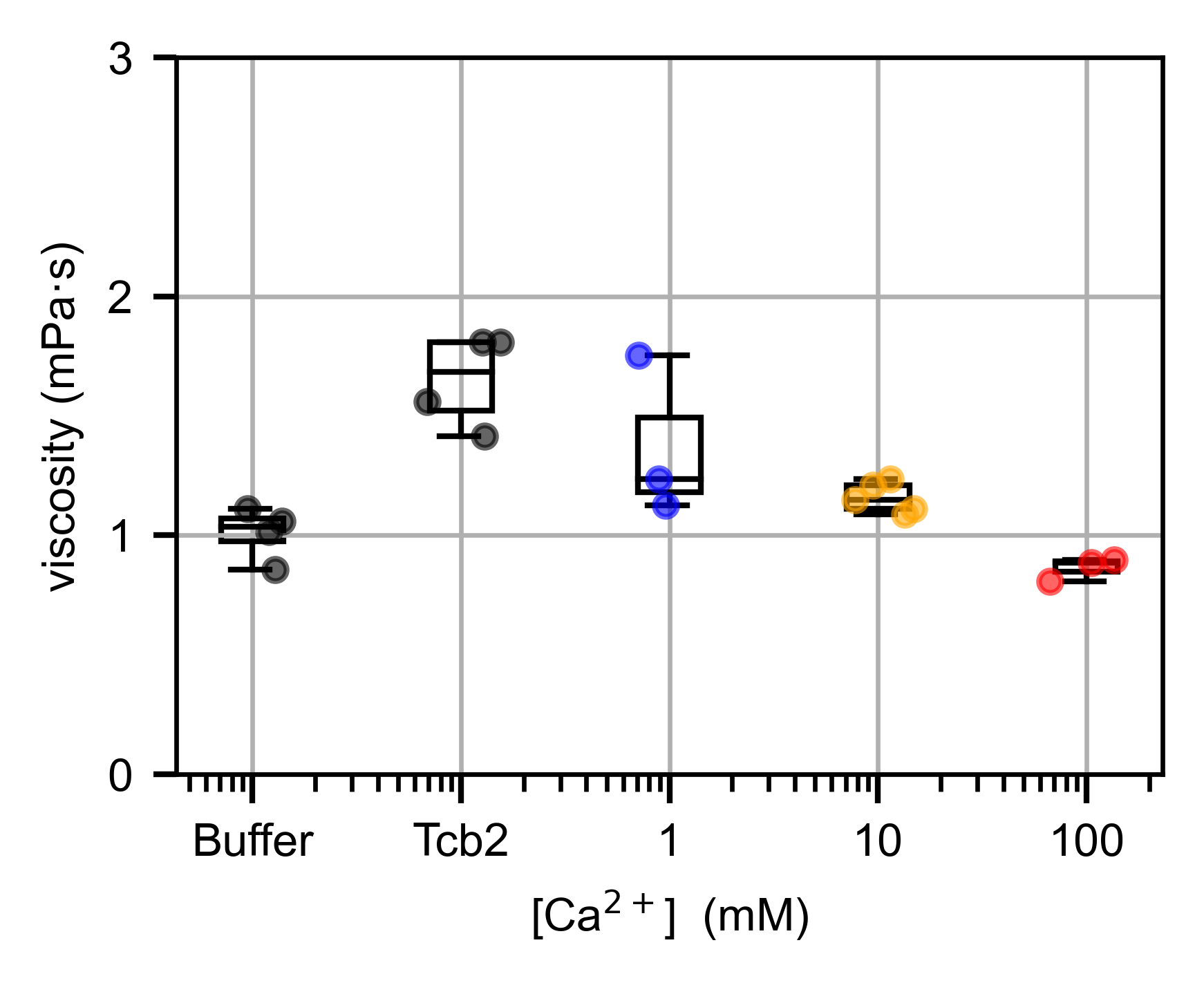}
    \caption{
    \textbf{Bulk viscosity of Tcb2 solutions at different Ca$^{2+}$ concentrations.}
    Each point is an independent measurement on a microfluidic rheometer (RheoSense). Pure water ($\eta = 1$\,cP) is shown for reference.
    }
    \label{SI:viscosity}
\end{figure}

Micro-viscometer measurements of the Tcb2 protein solution across different Ca$^{2+}$ concentrations are shown in Fig.~\ref{SI:viscosity}. These bulk measurements complement the microrheological data obtained with optical tweezers by providing an independent characterization of the viscous response. However, as discussed in the main text, bulk rheometry cannot resolve the local heterogeneity inherent to Ca$^{2+}$-assembled Tcb2 networks. We performed bulk viscosity measurements of Tcb2 solutions across the same Ca$^{2+}$ range. The bulk viscosity decrease mildly with Ca$^{2+}$, from $\sim$ 1.2 \,cP at 1\,mM to $\sim$ 0.9 \,cP at 100\,mM.

\section{Microrheology results}

\subsection{Displacement-dependent stiffness}

\begin{figure}[h!]
    \centering
    \includegraphics[width=1\linewidth]{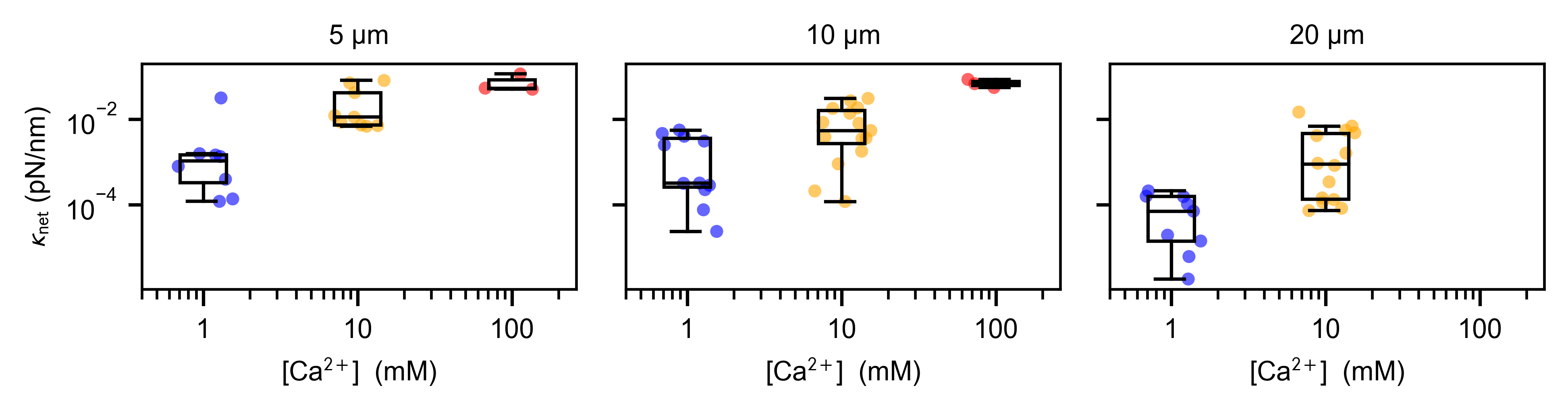}
    \caption{
    \textbf{Effective stiffness $\kappa_\mathrm{net}$ as a function of pull displacement at 1, 10, and 100\,mM Ca$^{2+}$ (interaction-zone values).}
    Each point is an independent bead. The 100\,mM / 20\,$\mu$m data is not accessible because the network resistance exceeds the trap's maximum holding force at this displacement (Fig.~\ref{SI:force_escape}).
    }
    \label{SI:displacement_moduli}
\end{figure}

To check whether the effective stiffness $k$ extracted in Fig.~2e of the main text falls within a linear-response regime, we measured $k$ across a range of pull displacements (5, 10, and 20\,$\mu$m) at all three Ca$^{2+}$ concentrations (Fig.~\ref{SI:displacement_moduli}). Within each Ca$^{2+}$ condition, $k$ remains approximately constant across the displacement range tested, indicating that the network responds linearly within the displacement range tested.

\subsection{Velocity-independence of $k$ across Ca$^{2+}$ conditions}

\begin{figure}[h!]
\centering
\includegraphics[width=0.5\linewidth]{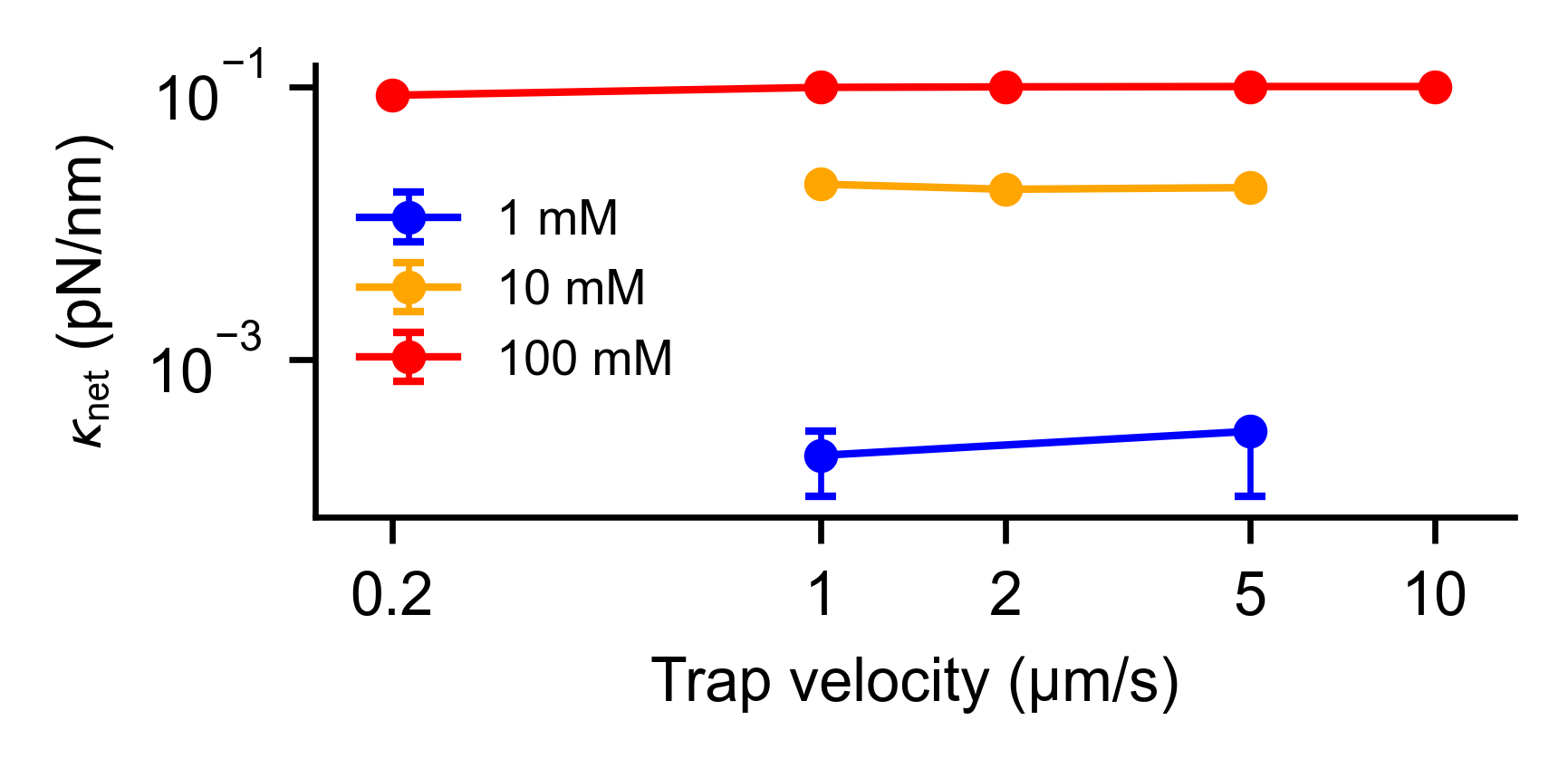}
\caption{
\textbf{Effective stiffness $\kappa_\mathrm{net}$ as a function of trap velocity at 1, 10, and 100\,mM Ca$^{2+}$.} At 100\,mM: 0.2, 1, 2, 5
and 10\,$\mu$m/s ($n = 6$, 12, 12, 12, 6). At 10\,mM: 1, 2 and
5\,$\mu$m/s ($n = 7$, 7, 10). At 1\,mM: 1 and 5\,$\mu$m/s ($n = 6$, 49).
}
\label{SI:kappa-vs-v}
\end{figure}

We drove the trap through forward and backward test and extracted $\kappa_\mathrm{net}$. Velocity coverage differs by condition: 100\,mM, three (1, 2, 5\,$\mu$m/s) at 10\,mM, and two (1, 5\,$\mu$m/s) at 1\,mM. Across the range tested, $k$ varies by no more than $1.16\times$ at 100\,mM and $1.09\times$ at 10\,mM (Fig.~\ref{SI:kappa-vs-v}), indicating that the slope-based stiffness is insensitive to moving speed. At 1\,mM the stiffness is larger ($1.5\times$).

\subsection{Force-displacement at 100\,mM Ca$^{2+}$}

\begin{figure*}[h!]
\centering
    \includegraphics[width=0.6\textwidth]{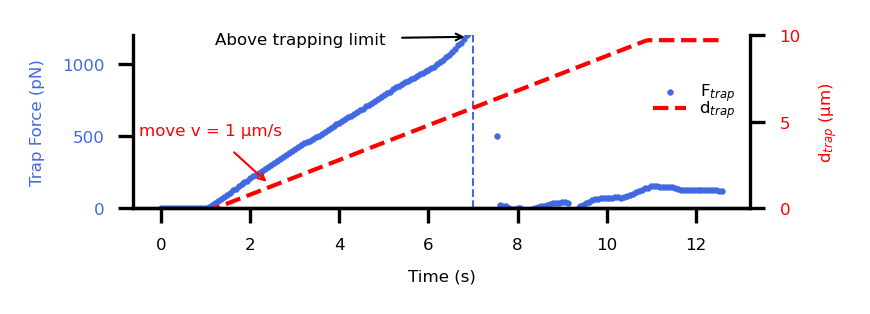}
    \caption{Force--displacement curve at 100\,mM Ca$^{2+}$. The force reaches $\sim$1000\,pN at $\sim$7\,s, exceeding the trap's maximum holding force and causing the bead to escape.}
    \label{SI:force_escape}
\end{figure*}

Figure~\ref{SI:force_escape} shows a representative force--displacement trace at 100\,mM Ca$^{2+}$. The force rises steeply with displacement and reaches the trap's maximum holding force ($\sim$1000\,pN) at $t \approx 7$\,s, at which point the bead escapes the trap.

\subsection{Conversion of effective stiffness to a shear modulus}

The effective stiffness $k$ reported in Fig.~2e of the main text characterises the
bead--network system rather than the material alone, and so cannot be compared
directly with moduli reported for other biopolymer networks. For a spherical probe
of radius $a$ embedded in a homogeneous elastic medium, the two are related through
the generalized Stokes--Einstein relation~\cite{Robertson-Anderson2018-gn,Li2023-rv},

\begin{equation}
G = \frac{k}{6 \pi a},
\label{eq:gse}
\end{equation}

which places our measurements on the same scale as bulk rheology. With the
4.0--4.9\,$\mu$m diameter beads used here ($a \approx 2.2$\,$\mu$m), Eq.~\ref{eq:gse}
gives the values in Table~\ref{SI:modulus}.

\begin{table}[h!]
\centering
\renewcommand{\arraystretch}{1.2}
\begin{tabular}{l c c c}
\hline
\textbf{[Ca$^{2+}$]} & $\boldsymbol{n}$ & $\boldsymbol{\kappa_\mathrm{net}}$ \textbf{(pN/nm)} & $\boldsymbol{G}$ \textbf{(Pa)} \\
\hline
1\,mM   & 29 & $(1.0 \pm 0.3)\times10^{-3}$ & $0.024 \pm 0.007$ \\
10\,mM  & 38 & $(8.1 \pm 1.4)\times10^{-3}$ & $0.19 \pm 0.03$ \\
100\,mM &  6 & $(7.2 \pm 1.0)\times10^{-2}$ & $1.7 \pm 0.2$ \\
\hline
\end{tabular}
\caption{Effective stiffness $\kappa_\mathrm{net}$ and the corresponding shear
modulus $G$ obtained from Eq.~\ref{eq:gse}. Values are mean $\pm$ standard error
across all force-displacement measurements (Fig.~2e of the main text).}
\label{SI:modulus}
\end{table}

At 1\,mM Ca$^{2+}$ the modulus $G \approx
0.024$\,Pa is  well below that of an entangled actin network
($G' \sim 0.01$--$0.1$\,Pa~\cite{Gardel2003-gl}), consistent with the
viscosity-dominated response seen in the recoil assays. At 10\,mM, $G \approx
0.19$\,Pa falls within the range reported for crosslinked actin
($0.1$--$1$\,Pa~\cite{Tharmann2007-yz}), and at 100\,mM, $G \approx 1.7 $\,Pa exceeds
it. A single-component, motor-free protein network, therefore, reaches stiffness
comparable to and beyond that of crosslinked actin purely through changes in Ca$^{2+}$
concentration.

We note that Eq.~\ref{eq:gse} assumes a homogeneous, incompressible medium and no-slip coupling at the bead surface. Ca$^{2+}$-assembled Tcb2 networks are
heterogeneous at 100~mM Ca$^{2+}$, so the values in
Table~\ref{SI:modulus} should be read as effective moduli averaged over the
bead surface rather than as exact continuum properties; the uncertainty in bead
radius alone ($a = 2.0$--$2.45$\,$\mu$m) contributes a  10\% deviation.

\section{Hysteresis under repeated back-and-forth pulls}
\label{SI:hysteresis}

We translated the trap back and forth at constant speed, so each cycle gives an outward and a return stroke over the same displacements. The trap oscillates
about its starting point, so successive cycles pull the bead in opposite directions and the force alternates sign. Here we plot the absolute displacement and
flip the sign of the force in the cycles pulled in the negative direction.

\begin{figure}[h!]
\centering
\includegraphics[width=0.48\linewidth]{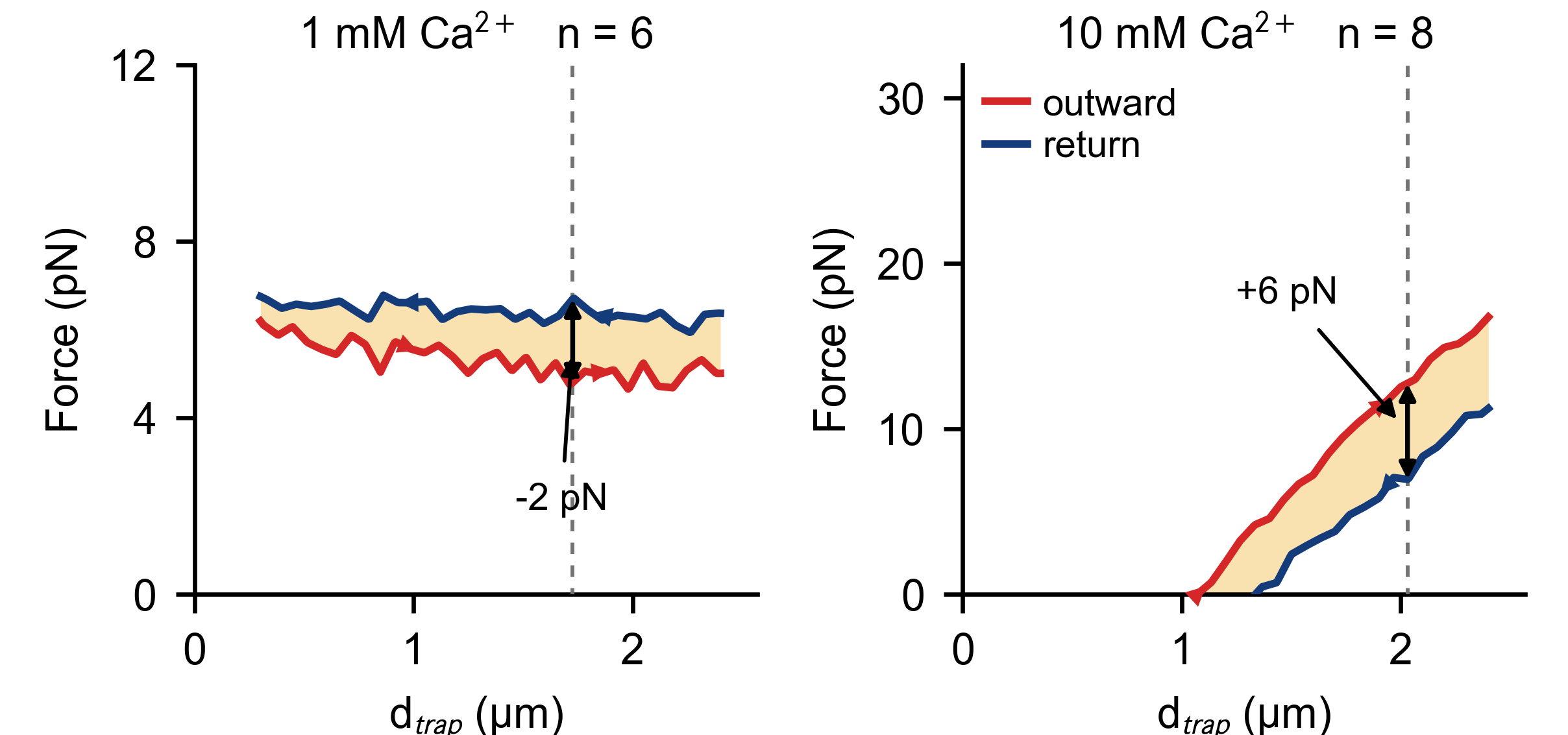}
\caption{Hysteresis at $1$ and $10$~mM Ca$^{2+}$. Outward (red) and return
(blue) branches of a single out-and-back cycle, each panel one bead, out of
$n$ cycles recorded. Shading marks the area between branches, the arrow the
peak separation. Fig.~2f shows the same measurement at $100$~mM.}
\label{SI:hysteresis_fig}
\end{figure}

Figure~\ref{SI:hysteresis_fig} shows a single out-and-back cycle from each run,
chosen so that half the cycles of that run show a larger offset and half a smaller one. The arrow marks the largest separation reached in that cycle, $-2$\,pN at $1$\,mM and $+6$\,pN at $10$\,mM. Averaged over all cycles of the same bead, the offset is $0.5 \pm 4.0$\,pN at $1$\,mM and $4.0 \pm 1.4$\,pN at $10$\,mM. We repeated the measurement on three independent beads at $100$\,mM and on two at $1$ and $10$\,mM.

\section{Recoil analysis}

\subsection{Recoil statistics}

\begin{figure}[h!]
\centering
\includegraphics[width=0.3\linewidth]{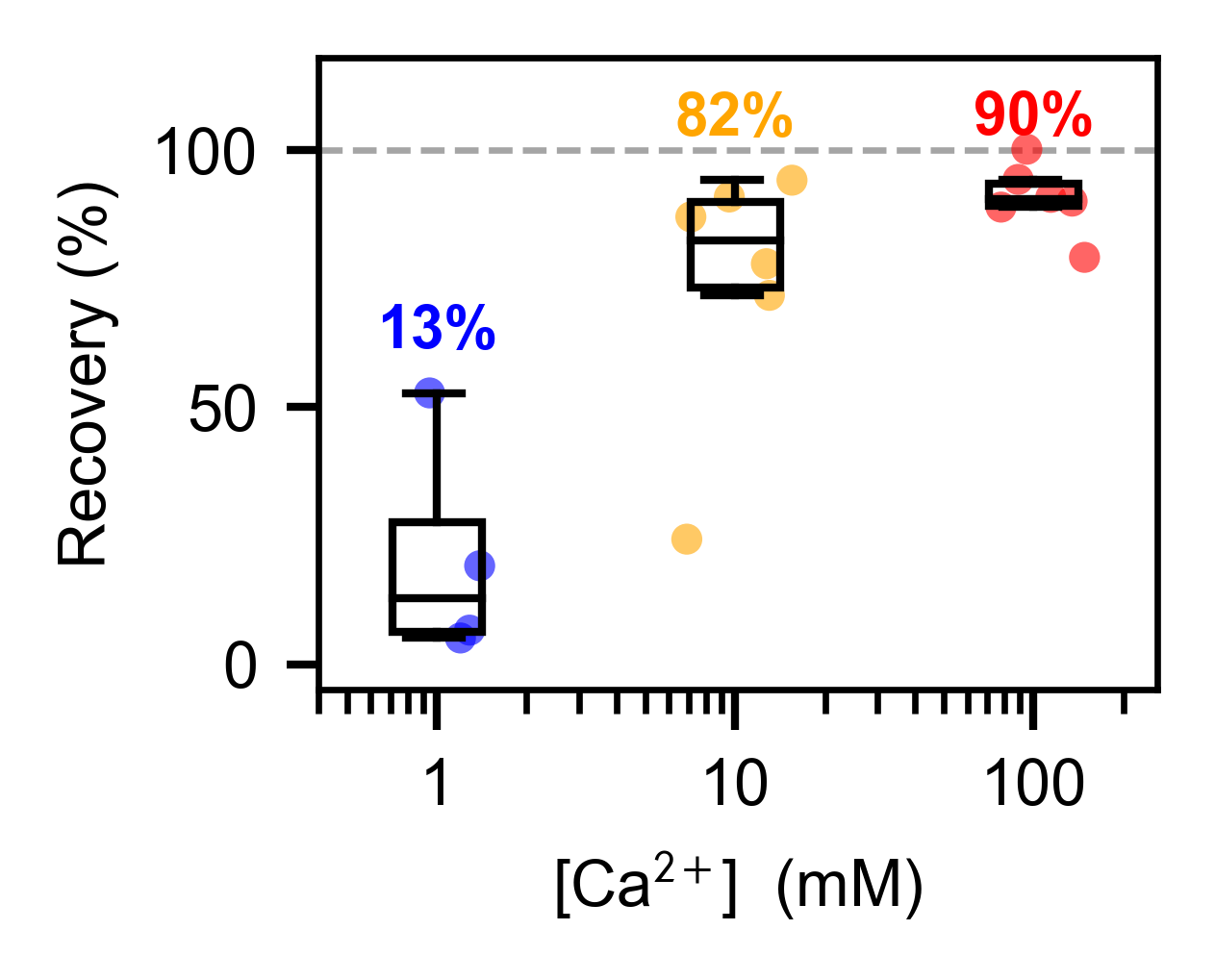}
\caption{
\textbf{Recovery fraction at different Ca$^{2+}$ concentrations.}
Recovery fraction $R = (D_\mathrm{final} - D_\mathrm{release})/(D_\mathrm{initial} - D_\mathrm{release})$ at 1, 10, and 100\,mM Ca$^{2+}$. Each point is an independent force--displacement measurement.}
\label{SI:recoil}
\end{figure}

We quantified the recoil response from the single-direction experiments in Fig.~3 of the main text. The recoil rate increases systematically with Ca$^{2+}$ (Fig.~\ref{SI:recoil}): 
median recovery of 13\% at 1\,mM ($n = 4$), 82\% at 10\,mM ($n = 6$), and 90\% at 100\,mM ($n = 6$).
\section{Background and Methods}

\subsection{Comparison with other biomolecular systems}

Table~\ref{SI:subcellular_mechanics} summarizes representative viscoelastic properties of DNA, RNA, and cytoskeletal components, providing context for the stiffness and force scales measured in Tcb2 networks. The closest motor-free analog actin networks exhibits storage moduli of 0.01--1\,Pa~\cite{Gardel2003-gl,Tharmann2007-yz}, while single motor proteins span a wider range, from $\sim$0.05~pN/nm for dynein~\cite{Kinoshita2018-ap} to $\sim$1.8~pN/nm for myosin~II~\cite{Lewalle2008-mx}. The Tcb2 stiffness reported here spans both regimes as Ca$^{2+}$ is varied.

\setlength{\tabcolsep}{4pt}
\renewcommand{\arraystretch}{1.2}
\begin{table}[h!]
    \centering
    \begin{tabular}{|p{3cm}|p{2.5cm}|p{2.8cm}|p{2.5cm}|p{1.5cm}|p{2.5cm}|p{1.5cm}|}
        \hline
        \textbf{Component} & \textbf{Region/Form} & \textbf{Property} & \textbf{Value $\pm$ Range} & \textbf{Units} & \textbf{Method} & \textbf{Ref.} \\
        \hline
        DNA (single molecule) & Double helix & Stretch modulus & 1000--1200 & pN & Optical tweezers & \cite{Wang1997-pl} \\
        RNA (double strand) & dsRNA & Stretch modulus & 350--600 & pN & Magnetic tweezers & \cite{Herrero-Galan2013-ag} \\
        Chromatin fiber & --- & Stretch modulus & 5 & pN & Optical tweezers & \cite{Cui2000-cn} \\
        Chromatin fiber & Reconstituted & Stretch modulus & 150 & pN & Optical tweezers & \cite{Bennink2001-yb} \\
        Actin network & --- & Storage modulus ($G'$) & 0.01--0.1 & Pa & Microrheology & \cite{Gardel2003-gl} \\
        Actin network & crosslinked & Storage modulus ($G'$) & 0.1--1 & Pa & Microrheology & \cite{Tharmann2007-yz} \\
        Actin filament & Single filament & Young's modulus ($E$) & 310 & MPa & Optical tweezers & \cite{Kikumoto2006-th} \\
        Microtubule & Paclitaxel-free & Young's modulus ($E$) & 460 & MPa & Optical tweezers & \cite{Kikumoto2006-th} \\
        Microtubule & Paclitaxel-stabilized & Young's modulus ($E$) & 120 & MPa & Optical tweezers & \cite{Kikumoto2006-th} \\
        Myosin~II & --- & Stiffness & $1.79 \pm 0.7$ & pN/nm & Optical tweezers & \cite{Lewalle2008-mx} \\
        Dynein & Single motor & Binding stiffness & 0.052 & pN/nm & Optical tweezers & \cite{Kinoshita2018-ap} \\
        \hline
    \end{tabular}
    \caption{Viscoelastic properties of DNA, RNA, and cytoskeletal components measured using optical tweezers and other biophysical methods.}
    \label{SI:subcellular_mechanics}
\end{table}

\subsection{Tcb2 plasmid and sequence}
\begin{figure}[h!]
    \centering
    \includegraphics[width=0.5\linewidth]{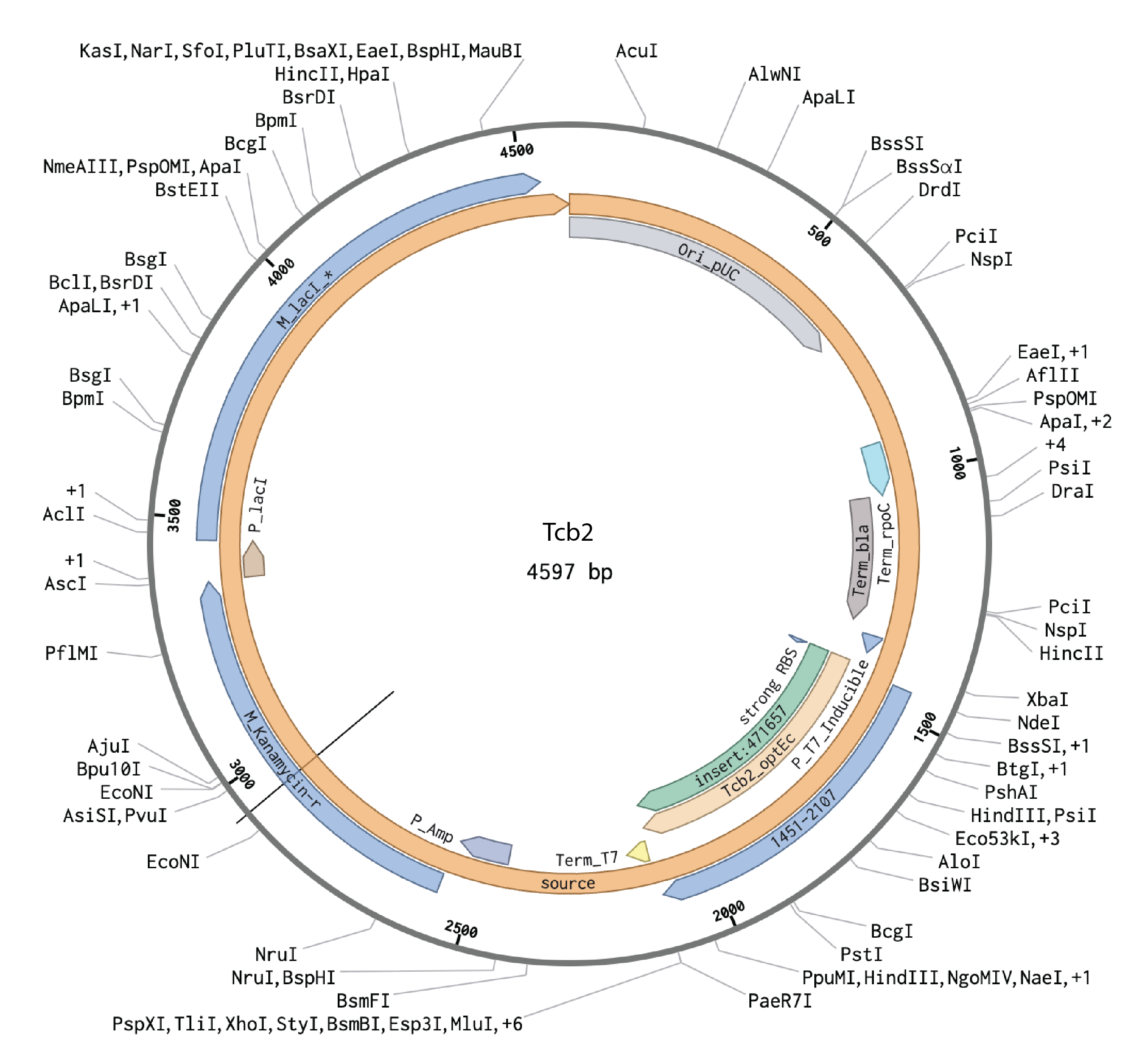}
    \caption{Plasmid map of Tcb2 with T7 promoter and kanamycin resistance.}
    \label{SI:plasmid}
\end{figure}

The protein sequence is reported in UniProtKB P09226~\cite{ncbi-Tcb2}. The Tcb2 gene was codon-optimized for \textit{E.~coli} and cloned into the pJ411 vector (4{,}597\,bp) under a T7 promoter with LacI regulation (Fig.~\ref{SI:plasmid}). The vector uses kanamycin for selection. Expression and purification followed our previous protocol~\cite{Lei2025-dq}.

\begin{align*}
 1   & \quad \text{MAQYSQTLRSSGFTSTVGLTDIEGAKTVARRIFENYDKGRKGRIENTDCVPMITEAYKSFNSFFAPSSDD}\quad 69 \\
 70  & \quad \text{IKAYHRVLDRNGDGIVTYQDIEELCIRYLTGTTVQRTIVTEEKVKKSSKPKYNPEVEAKLDVARRLFKRY}\quad 139 \\
 140 & \quad \text{DKDGSGQLQDDEIAGLLKDTYAEMGMSNFTPTKEDVKIWLQMADTNSDGSVSLEEYEDLIIKSLQKAGIR}\quad 209 \\
 210 & \quad \text{VEKQSLVF} \quad218 \\
\end{align*}

\section*{Supplementary Movie}

\noindent\textbf{Movie S1.} \textbf{Confocal imaging and optical-tweezer recoil of reconstituted Tcb2 networks at 1, 10, and 100\,mM Ca$^{2+}$.}

\vspace{0.5em}

\noindent\textit{Section I: Confocal imaging.} 3D confocal reconstructions of Tcb2 networks at 1, 10, and 100\,mM Ca$^{2+}$ (interaction-zone values). Scale bars: 50\,$\mu$m.

\vspace{0.5em}

\noindent\textit{Section II: Recoil experiments.} Bright-field recordings of optically trapped beads in Tcb2 networks. Single-direction recoil at 1, 10, and 100\,mM Ca$^{2+}$ after a 10\,$\mu$m, 1\,$\mu$m/s pull, followed by sequential two-dimensional L-shaped recoil at 100\,mM Ca$^{2+}$.

\bibliographystyle{unsrt}
\bibliography{proteinnetwork}